\documentclass[twocolumn,floatfix, showpacs]{revtex4}

\usepackage[final]{epsfig}
\usepackage{amsmath}
\begin{document}
 
\title{Using Hard Dihadron Correlations to constrain Elastic Energy Loss}
 
\author{Thorsten Renk}
\email{thorsten.i.renk@jyu.fi}
\affiliation{Department of Physics, P.O. Box 35, FI-40014 University of Jyv\"askyl\"a, Finland}
\affiliation{Helsinki Institute of Physics, P.O. Box 64, FI-00014 University of Helsinki, Finland}


\begin{abstract}
One contribution to the energy loss of hard partons propagating through a medium as created in ultrarelativistic heavy-ion (A-A) collisions are elastic scattering processes with medium constituents. The magnitude of this energy loss contribution depends crucially on the effective mass of the medium constituents  --- in the limit of non-recoiling static scattering centers, elastic energy loss vanishes. Thus, it is important to constrain the amount of elastic energy loss in order to gain information about the nature of the degrees of freedom in the medium as resolved by a hard parton. So far, the relative fraction of elastic (or rather incoherent) energy loss has been constrained from above by using pathlength dependent observables. However, using the observation that subleading gluon energy  dissipation into the medium is probed by some observables, such as hard dihadron correlations or the dijet asymmetry, a constraint from below can also be found. In this work, this idea is worked out within the in-medium shower evolution Monte Carlo (MC) code YaJEM-D.
\end{abstract}
 
\maketitle

\section{Introduction}

The energy loss of hard partons propagating through the soft medium created in heavy-ion collisions, leading to a suppression of high $P_T$ hadron spectra, has long been regarded as a promising tool to gain information on medium properties \cite{Jet1,Jet2,Jet3,Jet4,Jet5,Jet6}. However, there is a persistent puzzle in the literature with regard to the balance between radiative energy loss (where energy is carried by medium-induced gluon radiation) and elastic energy loss (where energy is carried in the recoil of medium scattering centers). 

If one starts with a first principles calculation from perturbative Quantum Chromodynamics (pQCD), modelling the medium as thermal distributions of quasi-free quarks and gluons with thermal masses of order $gT$ (where $g$ is the strong coupling with $\alpha_s = g^2/(4\pi)$ and $T$ the medium temperature) as motivated by thermal field theory, one naturally arrives at the conclusion that elastic energy loss is large \cite{Mustafa,Mustafa2,DuttMazumder,Djordjevic,Wicks,Ruppert,ElasticMC,Schenke}, possibly capable of accounting for about half of the energy carried away from a light quark passing through a medium.

If, on the other hand, one starts with the observation that elastic energy loss is an incoherent process and hence has a linear pathlength dependence in a constant medium, one can analyze the experimentally measured pathlength dependence by comparing models with the nuclear suppression factor $R_{AA}$ in non-central collisions as a function of the angle $\phi$ with the reaction plane. Such comparisons, both for a generic phenomenological model \cite{ElasticPhenomenology} and a detailed Monte-Carlo (MC) model \cite{ElasticMCRP} find that incoherent contributions to the total energy loss must be small and of the order of about 10\%. To the degree that elastic energy loss is only a part of all incoherent processes, this severely constrains the amount of elastic energy loss taking place in nature.

The inevitable conclusion is that the main assumption made in pQCD calculations of elastic energy loss, i.e. that the medium DOF are almost free (quasi)-particles which can take a sizeable amount of recoil energy away from a leading parton does not appear to be true in nature. In this case, constraining the amount of elastic energy loss offers information as to the nature of the scattering centers in the medium.

In this paper, we investigate a way to constrain elastic energy loss different from pathlength dependence by studying the interaction of subleading jet fragments. The mechanism is the same which has been invoked to explain the dijet asymmetry observed by ATLAS and CMS \cite{ATLAS,CMS} --- low $p_T$ shower partons are likely to be scattered out of a jet cone, and hence the combination of radiative energy loss as a source of soft gluons and elastic interactions decorrelating those from the jet can plausibly account for the observed dijet asymmetry \cite{Collimation}.   Here, we investigate the same effect in the context of triggered back-to-back correlations, which offer a somewhat less complicated environment free from the problems of jet reconstruction in a medium \cite{JetMedium}.

\section{The model}

\subsection{General framework}

We compute the strength of back-to-back hadron correlations by combining a MC pQCD calculation of back-to-back parton production \cite{IAA_old, IAA_new} with (geometry-dependent) medium-modified fragmentation functions obtained with the MC code YaJEM \cite{YaJEM1,YaJEM2} with the minimum virtuality scale down to which the shower is evolved in the medium determined by the prescription outlined in \cite{YaJEM-D} (YaJEM-D). Note that our modelling is constrained by multiple observables, among them the reaction plane angle dependence of  the nuclear suppression factor $R_{AA}$ for non-central collisions \cite{YaJEM-D} at RHIC, the reaction plane dependence of the dihadron correlation suppression factor $I_{AA}$ \cite{IAA_new} at RHIC and the nuclear suppression at LHC \cite{RAA_LHC}, thus we are only allowed to modify our modelling in a way that does not destroy agreement with these observables.

In LO pQCD, the production of two hard partons $k,l$ 
is described by
\begin{equation}
\label{E-2Parton}
  \frac{d\sigma^{AB\rightarrow kl +X}}{dp_T^2 dy_1 dy_2} \negthickspace 
  = \sum_{ij} x_1 f_{i/A}(x_1, Q^2) x_2 f_{j/B} (x_2,Q^2) 
    \frac{d\hat{\sigma}^{ij\rightarrow kl}}{d\hat{t}}
\end{equation}
where $A$ and $B$ stand for the colliding objects (protons or nuclei) and 
$y_{1(2)}$ is the rapidity of parton $k(l)$. The distribution function of 
a parton type $i$ in $A$ at a momentum fraction $x_1$ and a factorization 
scale $Q \sim p_T$ is $f_{i/A}(x_1, Q^2)$. The distribution functions are 
different for free protons \cite{CTEQ1,CTEQ2} and nucleons in nuclei 
\cite{NPDF,EKS98,EPS09}. The fractional momenta of the colliding partons $i$, 
$j$ are given by $ x_{1,2} = \frac{p_T}{\sqrt{s}} \left(\exp[\pm y_1] 
+ \exp[\pm y_2] \right)$.
Expressions for the pQCD subprocesses $\frac{d\hat{\sigma}^{ij\rightarrow kl}}{d\hat{t}}(\hat{s}, 
\hat{t},\hat{u})$ as a function of the parton Mandelstam variables $\hat{s}, \hat{t}$ and $\hat{u}$ 
can be found e.g. in \cite{pQCD-Xsec}. 

To account for various effects, including higher order pQCD radiation, transverse motion of partons in the nucleon (nuclear) wave function and effectively also the fact that hadronization is not a collinear process, the distribution is commonly folded with an intrinsic transverse momentum $k_T$ with a Gaussian distribution, thus creating a momentum imbalance between the two partons as ${\bf p_{T_1}} + {\bf p_{T_2}} = {\bf k_T}$.

Eq.~(\ref{E-2Parton}) is evaluated at midrapidity $y_1 = y_2 = 0$ and sampled using a MC code introduced in \cite{IAA_old} by first generating the momentum scale of the pair and then the (momentum-dependent) identity of the partons. A randomly chosen $k_T$ with a Gaussian distribution of width 2.5 GeV is then added to the pair momentum. 

Under the assumption that the distribution of vertices follows binary collision scaling as appropriate for a LO pQCD calculation, the probability density to find a vertex in the transverse plane is

\begin{equation}
\label{E-Profile}
P(x_0,y_0) = \frac{T_{A}({\bf r_0 + b/2}) T_A(\bf r_0 - b/2)}{T_{AA}({\bf b})},
\end{equation}
where the thickness function is given in terms of Woods-Saxon distributions of the the nuclear density
$\rho_{A}({\bf r},z)$ as $T_{A}({\bf r})=\int dz \rho_{A}({\bf r},z)$ and $T_{AA}({\bf b})$ is the standard nuclear overlap function $T_{AA}({\bf b}) = \int d^2 {\bf s}\, T_A({\bf s}) T_A({\bf s}-{\bf b})$ for impact parameter ${\bf b}$. Each parton pair is placed at a probabilistically chosen vertex $(x_0,y_0)$ sampled from this distribution with a random orientation $\phi$ with respect to the reaction plane.

Both partons are then propagated on eikonal paths through a hydrodynamical medium \cite{hydro3d} and for this path the leading and first two subleading fragments are computed using the medium-modified conditional probability densities $A_1(z_1, \mu)$, $A_2(z_1, z_2, \mu)$ and $A_3(z_1, z_2, z_3, \mu) \approx A_2(z_1+z_2, z_3, \mu)$ given the path \cite{IAA_LHC}. For this, we utilize YaJEM-D. Finally we check if there is a hadron in the event which fulfills the trigger condition. If not, we discard the event and start generating a new one. If there is a trigger hadron, we bin the remaining hadrons in the event on the near and away side in either $P_T^{assoc}$ or $z_T$. 

\subsection{Medium-modified fragmentation}

The key ingredient containing the information about the medium evolution and shower-medium interaction is the medium-modified fragmentation function (MMFF) which we compute in YaJEM-D.
The MC code YaJEM-D is based on the PYSHOW code \cite{PYSHOW} which is part of PYTHIA \cite{PYTHIA}. It simulates the evolution from a highly virtual initial parton to a shower of partons at lower virtuality in the presence of a medium down to a minimum scale $Q_0 = \sqrt{E/L}$ where $E$ is the energy of the shower initiator and $L$ is the in-medium pathlength. A detailed description of the model can be found in \cite{YaJEM1,YaJEM2,YaJEM-D}.

In the RAD (radiative energy loss) scenario, a  parton $a$ gains virtuality during its propagation time $\tau_a$ as

\begin{equation}
\label{E-Qgain}
\Delta Q_a^2 = \int_{\tau_a^0}^{\tau_a^0 + \tau_a} d\zeta \hat{q}(\zeta)
\end{equation}

which leads to extra branchings and soft gluon production where $\hat{q}(\zeta)$ is the local rate of virtuality gain. However, YaJEM-D also implements the so-called DRAG scenario in which propagating partons lose energy (and momentum) according to

\begin{equation}
\label{E-Drag}
\Delta E_a = \int_{\tau_a^0}^{\tau_a^0 + \tau_a} d\zeta D(\zeta)
\end{equation}

where $D(\zeta)$ is the local drag coefficient corresponding to a mean energy loss $dE/d\zeta$ per unit length. This model is also suited to treat incoherent, elastic energy loss. These coefficients are tied to the hydrodynamical parameters of the medium evolution model via

\begin{equation}
\label{E-qhat}
\hat{q}[D](\zeta) = K[K_D] \cdot 2 \cdot [\epsilon(\zeta)]^{3/4} (\cosh \rho(\zeta) - \sinh \rho(\zeta) \cos\psi)
\end{equation}

where $\psi$ is the angle between bulk medium flow and the parton direction, $\rho$ is the flow rapidity and $\epsilon$ is the medium energy density.  This expression is a proxy for the Lorentz-contracted density of scattering centers in the flowing medium as seen from a moving parton. $K$ and $K_D$ then parametrize the actual strength of the parton-medium interaction given this density.

In \cite{YaJEM2} it was found that both scenarios obey a scaling law, i.e. to a good approximation what matters for the end result are the integrals $\Delta E = \int d \zeta D(\zeta)$ and $\Delta Q^2 = \int d \zeta \hat{q}(\zeta)$ rather than the full functional dependence of the coefficients on $\zeta$. It was also found that both scenarios
give equally good descriptions of the nuclear suppression factor $R_{AA}$ at RHIC kinematics (their results do, however, differ for other more differential observables such as jet shapes \cite{YaJEM-Jet}). This implies that suitable combinations of $(\hat{q}, D)$ where one of the parameters does not vanish will also describe the $P_T$ dependence of $R_{AA}$. 

\begin{figure}[htb]
\begin{center}
\epsfig{file=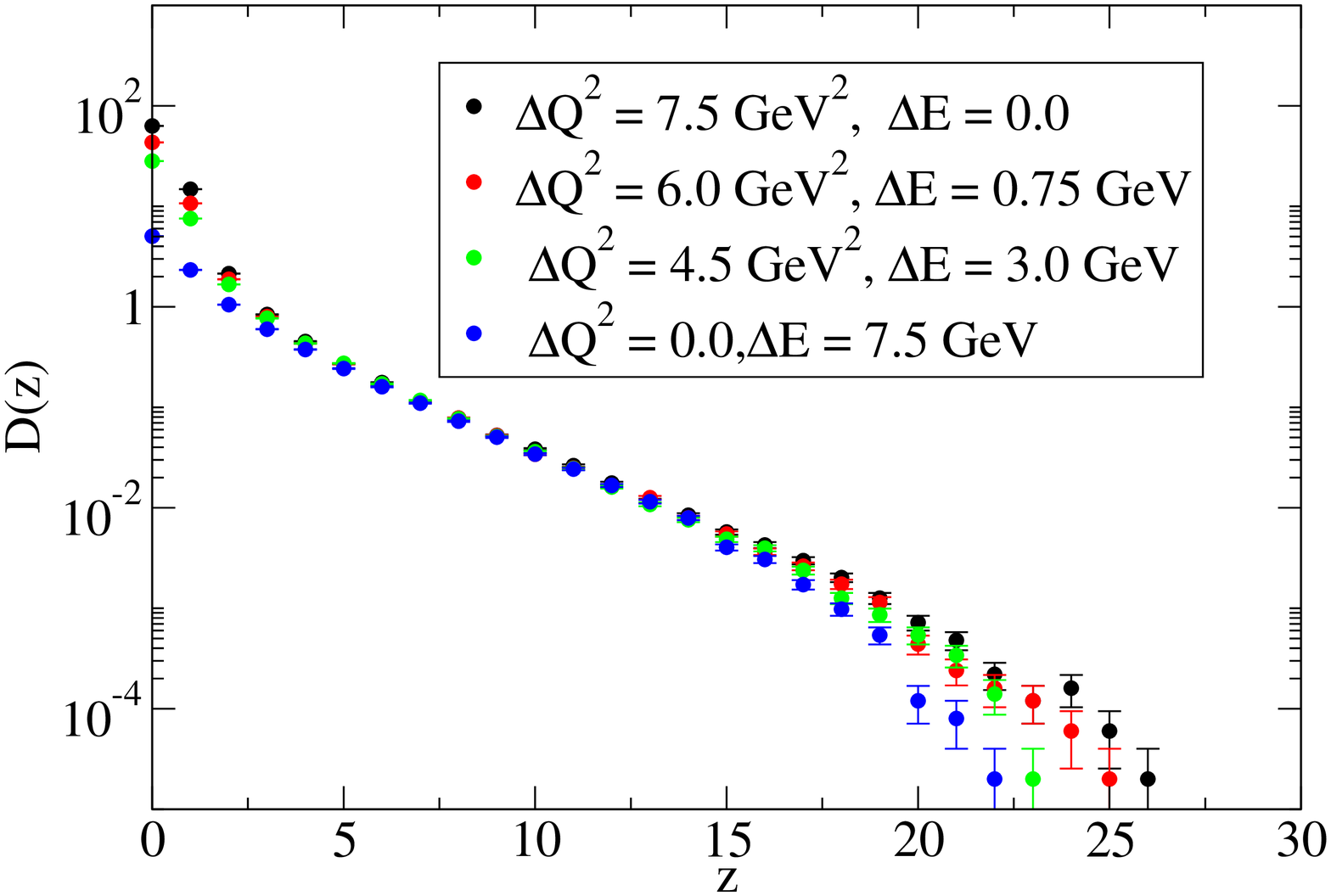, width=8.5cm}
\end{center}
\caption{\label{F-Dcomp}(Color online) A comparison of medium-modified fragmentation functions for a 30 GeV quark computed in YaJEM-D for various relative contributions of radiative and elastic energy loss, indicated by the mean accumulated virtuality $\Delta Q^2$ or the mean energy loss $\Delta E$ along the parton path.}
\end{figure} 

We can gain some insight into this by studying the MMFF for different parameter combinations  $(\hat{q}, D)$ (or rather their path-integrated values $(\Delta Q^2, \Delta E)$) in Fig.~\ref{F-Dcomp}. Folded with a pQCD parton spectrum, one finds that this fragmentation function has a mean $z$ of $\sim 0.3$, i.e. just where all scenarios agree. This explains why there is no pronounced difference between the scenarios for single inclusive hadron production in spite of the obvious differences in the MMFF at high and low $z$.

It is difficult to exploit the differences at high $z$ to determine which behaviour of the MMFF is realized in nature since high $z$ fragmentation is a rare process. However, the low $z$ behaviour is easily studied by observing subleading shower hadrons in back-to-back dihadron correlations on the away side. 

\section{Results}

\begin{figure}[htb]
\begin{center}
\epsfig{file=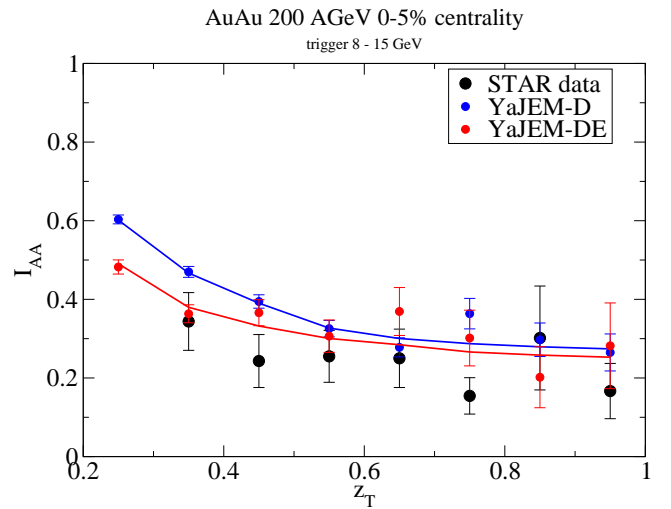, width=8.5cm}
\end{center}
\caption{\label{F-IAA-AuAu-away}(Color online) The away side $I_{AA}$ for RHIC  Au-Au collisions, computed using the in-medium shower MC code YaJEM-D with only radiative energy loss or with a 10\% contribution of elastic energy loss (YaJEM-DE) compared with STAR data \cite{STAR-DzT}.}
\end{figure}

We show the away side hard dihadron suppression factor $I_{AA}$ as a function of $z_T$ (momentum of observed away side hadron divided by trigger hadron momentum) computed as outlined above in Fig.~\ref{F-IAA-AuAu-away} and compare with STAR data \cite{STAR-DzT}. We do the comparison for two different scenarios: One (YaJEM-D) assumes pure radiative energy loss (and agrees with the result presented in \cite{IAA_new}), the other (YaJEM-DE) uses $\hat{q} = 0.8 \cdot \hat{q}_{max}$ and $D = 0.1 \cdot D_{max}$ (where $\hat{q}_{max}$ and $D_{max}$ are the parameter values one extracts for a fit to $R_{AA}$ assuming that the corresponding other parameter is zero, i.e. for a pure radiative or a pure elastic energy loss scenario). These values are motivated by the constraint that the elastic contribution should be no more than of order 10\% of the total energy loss as constrained by pathlength dependence. They also satisfy the constraint that the $P_T$ dependence of single hadron $R_{AA}$ should be unchanged within experimental errors.

We see that at high $z_T$ there is no statistically significant difference between the two scenarios. However, at low $z_T$ they clearly separate, in accordance with the behaviour seen in the MMFF in Fig.~\ref{F-Dcomp}, and the scenario including elastic energy loss describes the data much better in this region.

The physics interpretation of this result is simple: Medium-induced radiative energy loss creates a large number of soft gluons, which in turn leads not only to a depletion of the MMFF at high $z$ ('quenching') but also to a rapid growth at low $z$ and consequently of $I_{AA}^{away}$ at low $z_T$. A certain contribution of elastic energy loss is needed to dissipate the energy of these soft gluons into the medium and thus tame the growth of the MMFF. 

While it is difficult to extract relative fractions of energy loss in a medium-modified shower as such since it is impossible to tag any particular branching and assign it to either vacuum radiation or medium-induced radiation (see discussion in \cite{YaJEM2}), based on the magnitude of the transport coefficients compared to pure radiative or elastic scenarios, it seems that a fraction of about 10\% elastic contribution which is allowed by pathlength dependence is also sufficient to account for the observed modification of the fragmentation function in dihadron correlations. However, note that while it is true that choosing a value of 10\% $D_{max}$ corresponds to also 10\% \emph{mean} energy loss (the actual energy loss in any single MC event may differ) with respect to the value obtained with $D_{max}$, the same relation is not true for $\hat{q}$ which has a complicated non-linear relation with gluon radiation which includes LPM suppression of soft gluons during decoherence time and kinematical restrictions. Thus, ratios of the coefficients $D, \hat{q}$ do \emph{not} imply the same ratio of mean energy loss from the leading parton. Therefore, a more precise whay of thinking about the problem is fixing the relative strength of the parameters $(K,K_D)$ which determine the value of $(\hat{q}, D)$ given the hydrodynamical parameters of the medium evolution model.

Given that the same mechanism, i.e. dissipation of soft gluon energy into the medium, is used to explain the dijet asymmetry \cite{Collimation} observed by ATLAS and CMS \cite{ATLAS,CMS}, a crucial test for the combination of radiative and elastic energy loss determined in the scenario YaJEM-DE is whether or not the model is able to account for the measured dijet asymmetry. This question will be addressed in future work.

If the idea that elastic energy loss is observed dissipating the energy of low $z$ gluons turns out to be correct, pathlength dependence can be used to constrain the amount of incoherent energy loss from above whereas dissipation of energy in soft gluons can be used to constrain it from below, i.e. a relatively precise determination of incoherent energy loss should be viable. This in turn may greatly help to address the question what the microscopical degrees of freedom in the medium as probed by a hard parton are.

\begin{acknowledgments}
 
I would like to thank Steffen Bass and Chiho Nonaka who supplied the medium evolution model used in this work, and in addition Kari Eskola for valuable discussions. This work is supported by the Academy researcher program of the
Academy of Finland, Project No. 130472. 
 
\end{acknowledgments}

\end{document}